\documentclass[11pt,USletter]{article}
\usepackage[utf8]{inputenc}
\usepackage[english]{babel}
\usepackage{amsmath}
\usepackage{amsfonts}
\usepackage{amssymb}
\usepackage{graphicx}
\usepackage[left=2cm,right=2cm,top=2cm,bottom=2cm]{geometry}

\usepackage{cite}
\usepackage{amsmath,amssymb,amsfonts}
\usepackage{graphicx}
\usepackage{textcomp}

\usepackage{url}
\usepackage{hyperref}
\usepackage{cleveref}
\usepackage{float}
\usepackage{longtable}
\usepackage{supertabular}
\usepackage{caption}

\usepackage{forest}

\usepackage{xcolor}

\newcommand{\blue}[1]{{\color[RGB]{0,0,255} #1}}

\usepackage{indentfirst,latexsym,bm,bbm}
\usepackage{mathrsfs}
\usepackage{amsmath}
\usepackage{amssymb}
\usepackage{amsfonts}
\usepackage{amsthm}
\usepackage{mathbbol}

\newtheorem{thm}{Theorem}

\DeclareMathOperator{\dif}{d}         

\DeclareMathOperator{\Zern}{Z}    % Zernike
\DeclareMathOperator{\MOD}{mod}
\renewcommand{\mod}{\MOD}

\newcommand{\mfloor}[1]{ \left\lfloor {#1} \right\rfloor }
\newcommand{\mceil}[1]{ \left\lceil {#1} \right\rceil }
\newcommand{\mpair}[2]{ \left\langle {#1}, {#2} \right\rangle}

\newcommand{\mi} {\mathrm{i}}

\newcommand{\me} {\mathrm{e}}

\newcommand{\set}[1]{\left\{ #1 \right\}}
\newcommand{\seq}[1]{\langle #1 \rangle}
\newcommand{\abs}[1]{\left| #1 \right|}
\newcommand{\braket}[2]{ \langle #1 | #2 \rangle}

\newcommand{\card}[1]{\abs{#1}}

\newcommand{\Zernpoly}[2]{\operatorname{Z}_{#1}^{#2}}
\newcommand{\Radipoly}[2]{\operatorname{R}_{#1}^{#2}}

\newcommand{\scrd}[2]{{#1}_{\mathrm{#2}}}

\usepackage{algorithm}         
\usepackage{algorithmicx}
\usepackage{algpseudocode}
\usepackage{xcolor}

\newcommand{\Algr}{\textbf{Algorithm}~}
\newcommand{\Fig}{FIGURE~}
\newcommand{\Tab}{TABLE~}

\algnewcommand\algorithmicswitch{\textbf{switch}}
\algnewcommand\algorithmiccase{\textbf{case}}
\algnewcommand\algorithmicdefault{\textbf{default}}
\algnewcommand\algorithmicassert{\texttt{assert}}
\algnewcommand\Assert[1]{\State \algorithmicassert(#1)}%
\algdef{SE}[SWITCH]{Switch}{EndSwitch}[1]{\algorithmicswitch\ #1\ \algorithmicdo}{\algorithmicend\ \algorithmicswitch}%
\algdef{SE}[CASE]{Case}{EndCase}[1]{\algorithmiccase\ #1}{\algorithmicend\ \algorithmiccase}%
\algdef{SE}[DEFAULT]{Default}{EndDefault}[1]{\algorithmicdefault\ #1}{\algorithmicend\ \algorithmicdefault}%

\makeatletter
\renewcommand{\ALG@name}{Algorithm}
\newenvironment{breakablealgorithm}
{% begin of the breakablealgorithm
	\begin{center}
		\refstepcounter{algorithm}% New algorithm
		\setlength{\baselineskip}{15pt} 
		\renewcommand{\caption}[2][\relax]{% Make a new \caption
			\hrule height.9pt depth0pt \kern3pt
			{\raggedright\textbf{\ALG@name~\thealgorithm} ##2\par}%
			\ifx\relax##1\relax % #1 is \relax
			\addcontentsline{loa}{algorithm}{
				\protect\numberline{\thealgorithm}##2}%
			\else % #1 is not \relax
			\addcontentsline{loa}{algorithm}{
				\protect\numberline{\thealgorithm}##1}%
			\fi
			\kern2pt\hrule\kern2pt
		}
	}{% end of the breakablealgorithm
		\kern3pt\hrule\relax% \@fs@post for \@fs@ruled 
	\end{center}
}

\usepackage{caption}
\usepackage{listings}
\newcommand{\cpvar}[1]{\texttt{#1}}

\newcommand{\ProcName}[1]{\textsc{#1}}

\title{\textbf{An Automatic Method for Generating Symbolic Expressions of Zernike Circular Polynomials}\thanks{This work was supported in part by the Hainan Provincial
Natural Science Foundation of China under Grant 
2019RC199, in part by the Hainan Provincial Education and Teaching Reform Project of Colleges and Universities under
Grant Hnjg2019-46,
and in part by the National Natural Science Foundation of China under Grant 62167003.}}

\author{Hong-Yan Zhang\thanks{Correspondence author, e-mail: hongyan@hainnu.edu.cn}, Yu Zhou and Fu-Yun Li\\
School of Information Science and Technology, Hainan Normal University\\ No. 99, Rd. Longkun South, Haikou 571158, P. R. China}
\date{Jan. 7, 2023}

\begin{document}
\maketitle

\begin{abstract}
Zernike circular polynomials (ZCP) play a significant role in optics engineering. The symbolic expressions for ZCP are valuable for theoretic analysis and engineering designs. However, there are still two problems which remain open: firstly, there is a lack of sufficient mathematical formulas of the ZCP for optics designers; secondly the formulas for inter-conversion of Noll's single index and Born-Wolf's double indices of ZCP are neither uniquely determinate nor satisfactory. An automatic method for generating symbolic expressions for ZCP is proposed based on five essential factors: the new theorems for converting the single/double indices of the ZCP, the robust and effective numeric algorithms for computing key parameters of ZCP, the symbolic algorithms for generating mathematical expressions of ZCP, and meta-programming \& \LaTeX{} programming for generating the table of ZCP. The theorems, method, algorithms and system architecture proposed are beneficial to both optics design process, optics software, computer-output typesetting in publishing industry as well as STEM education.    
\end{abstract}

\tableofcontents

\section{Introduction}
In optics engineering, the \textit{Zernike circular polynomials} (ZCP), also named Zernike polynomials for simplicity,  are essential for representing aberrations in imaging system, optics design and optical testing \cite{BornWolf-1999,Mahajan-1981,Jasssen-2010,Shakibaei-2013,Diaz-2013,Mathar-2008,Buhren2018,Berger2022-ZernikeAberr,OptikShopTest-2007}. Mathematically, ZCP are a sequence of bivariate polynomials defined on the unit disk 
\begin{align*}
\mathcal{D} &= \set{(x,y): 0\le x^2+y^2\le 1} \\
&= \set{(\rho,\theta): 0\le \rho\le 1, 0\le \theta\le 2\pi}
\end{align*} 
derived from the circular pupils of imaging system. Named after Frits Zernike, they play an important role in beam optics and optics design\cite{Zemax-2008}, atmospheric turbulence \cite{Noll-1976} and image processing \cite{Chong-2003}.
The ZCP are orthogonal functions which  represent balanced aberrations that yield minimum variance\cite{OptikShopTest-2007}. The computation of ZCP are significant for theoretic analysis and engineering applications. The numerical algorithms are discussed in lots of literature such as \cite{Shakibaei-2013,Kintner-1976,Chong-2003,BbtZernike2022}. However, the symbolic computation for automatically generating the mathematical expressions for the ZCP is missing. It may be inconvenient or troublesome for the optics designers to look for the formulas of ZCP with "high" orders. Actually, there are less $28\sim 40$ and $70$ expressions for the ZCP in ZEMAX and CodeV respectively.  

\begin{figure*}[htbp]
\centering
\includegraphics[width=0.8\textwidth]{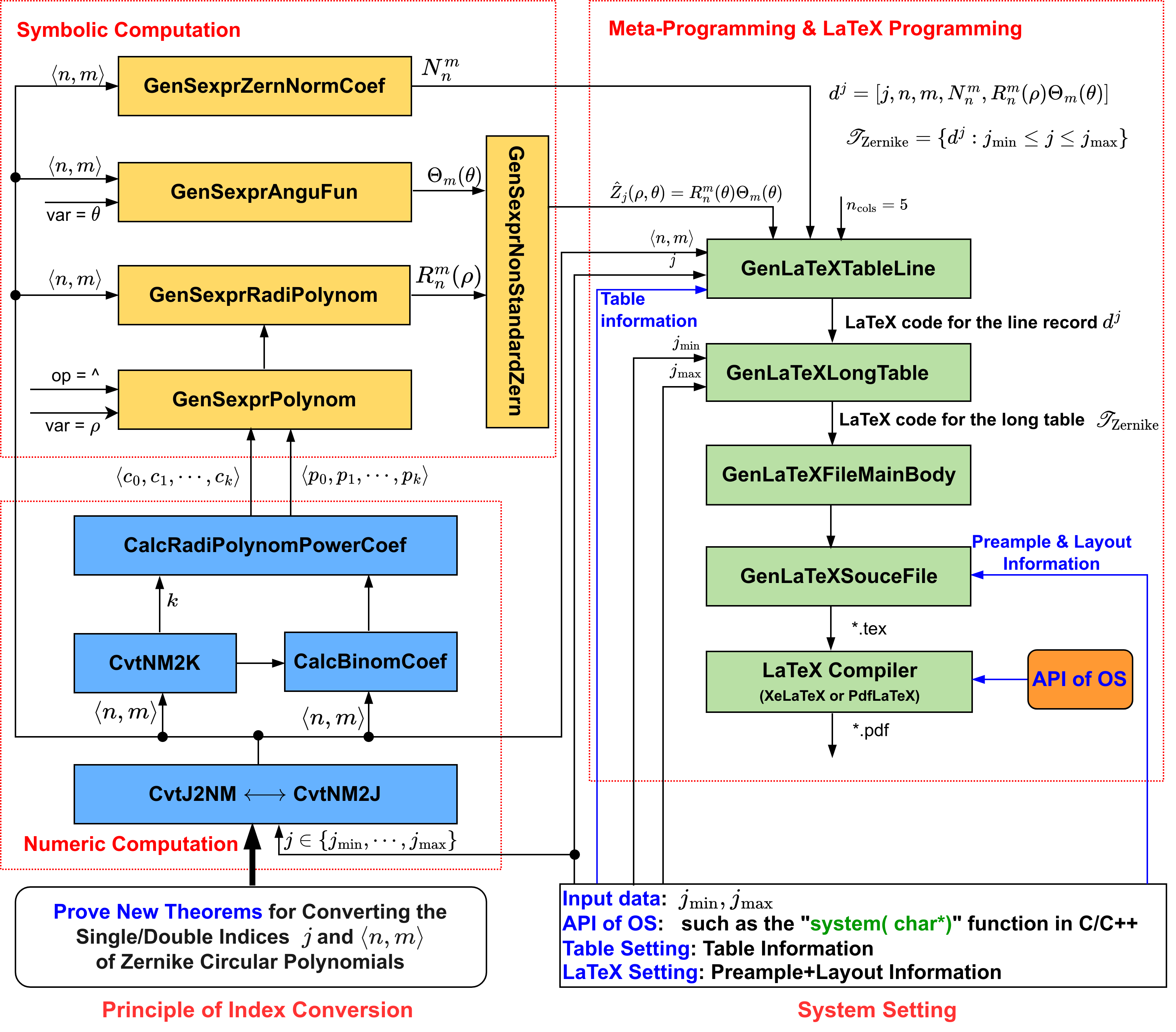} 
\caption{System architecture for automatically generating a long table of 
the mathematical expressions of ZCP $\set{\Zern_j(\rho,\theta): \scrd{j}{min}\le j \le \scrd{j}{max}}$}
\label{fig-system-frame}
\end{figure*}

The purpose of this paper is to propose an automatic method for generating mathematical expressions for the ZCP. \Fig 
\ref{fig-system-frame} illustrates the framework of our work. There are five modules for the system architecture:
\begin{itemize}
\item Principle of Index Conversion, which includes our new theorems and proofs about the single/double index for the ZCP;
\item Numeric computation, which computes key parameters robustly and effectively for large integers so as to avoid the overflow problem in computing factorials;
\item Symbolic computation, which deals with the algorithms for automatic generating symbolic expressions of the ZCP;
\item Meta-programming and \LaTeX{} programming, which generates the long table of ZCP and outputs the corresponding pdf file for printing, reading and reference;
\item System setting, which sets the necessary information for the mathematical formula of ZCP and the geometric configuration of the long table of mathematical expressions. 
\end{itemize}

The contents of this paper are organized as follows: 
Section \ref{sec-preliminary} deals with the preliminaries of backgrounds; Section 
\ref{sec-newprinciple} copes with the principle of inter-conversion of the single/double index for computing ZCP; Section
\ref{sec-gen-table} focuses on how to generate long table of mathematical expressions of ZCP; Section \ref{sec-conclusion} is the conclusion.

\section{Preliminaries} \label{sec-preliminary}

\subsection{Fundamentals of Mathematics}
 \subsubsection{Notations for Integers}

For an integer $n\in \mathbb{Z} = \set{0, \pm 1, \pm 2, \cdots}$, it is even if and only if $2 \mid n$, i.e., $2$ divides $n$ or equivalently  $n \equiv 0~(\mod 2)$; otherwise, it is odd if and only if $ 
2\nmid n$ or equivalently $n\equiv 1~(\mod 2)$. The sets of even and odd integers are denoted by 
$$
\mathbb{Z}_{\mathrm{odd}}  
= \set{i\in \mathbb{Z}: 2\nmid i}  
= \set{\pm 1, \pm 3, \pm 5, \cdots }
$$ 
and 
$$
\mathbb{Z}_{\mathrm{even}} 
= \set{i\in \mathbb{Z}: 2\mid i}
= \set{0, \pm 2, \pm 4, \pm 6, \cdots }
$$
respectively. The set of positive integers are $\mathbb{N} = \set{1, 2, 3, \cdots}$ and the set of non-negative integers is  
$\mathbb{Z}^+ = \mathbb{N}\cup \set{0} = \set{0, 1, 2, \cdots}$. Similarly, the set of negative integers is $\mathbb{Z}^{-} = -\mathbb{N} = \set{-1, -2, -3, \cdots}$. In consequence, the sets of non-negative even integers and non-negative odd integers can be denoted by  
$\scrd{\mathbb{Z}}{even}^+$  and $\scrd{\mathbb{Z}}{odd}^+$ respectively.

For any real number $x\in \mathbb{R}$, the unique integer $n = \mfloor{x}\in \mathbb{Z}$ such that $n \le x < n+1$ is called the floor of $x$. 
Similarly,  the integer $n'= \mceil{x} \in \mathbb{Z}$ such that $n'-1 < x \le n'$ is called the ceiling of $x$.

\subsubsection{Expression of General Polynomials}

A polynomial $T_n(x)$ with argument $x$ and degree $n$ can be written by
\begin{equation}
T_n(x) = \sum^n_{i=0} a_ix^{i} = a_0 + a_1x + \cdots + a_ix^{i} + \cdots + a_n x^{n}
\end{equation} 
If $a_i = 0$ for all odd $i$ , then $T_n(x)$ is an even polynomial. Similarly, if $a_i = 0$ for all even $i$, then $T_n(x)$ is an odd polynomial. 
Futhermore, some of the coefficients $a_0, a_1, \cdots, a_n$ may be missing if they equal zero. In consequence, we can denote a polynomial as
\begin{equation}
T(x) = \sum^k_{s=0}c_s x^{p_s} , \quad p_s\in \mathbb{Z}^+
\end{equation}
where $p_s\in \mathbb{Z}^+$. In other words, a polynomial can be described by the sequence of coefficients $\cpvar{c} = \seq{c_0, \cdots, c_k}$ and the corresponding sequence of power indices $\cpvar{p} = \seq{p_0, \cdots, p_k}$.

\subsubsection{Binomial and Tri-nomial Coefficients}
In general, for any real number $\alpha\in \mathbb{R}$ and non-negative $i\in \mathbb{Z}^+ = \set{0, 1, 2, \cdots}$ the generalized binomial expansion can be expressed by
$$
(1+x)^\alpha = \sum^\infty_{i=0}\binom{\alpha}{i}x^i, \quad x\in \mathrm{ROC} = (-1, 1)
$$
in which ROC is the region of convergence and
\begin{equation} 
\binom{\alpha}{i} = \frac{\alpha(\alpha-1)\cdots (\alpha-i+1)}{i!} 
\end{equation}
is the binomial coefficient \cite{TAOCP-1,GrahamKnuth-1994,SJZhang1996}.
Particularly, if $\alpha\in \mathbb{Z}^+$ is a positive integer, we have 
\begin{equation}
\binom{\alpha}{i} = \frac{\alpha!}{i!(\alpha-i)!}.
\end{equation} 
For any $r\in \mathbb{Z}^+$ we have
$$
(x_1 + x_2 +  x_3)^r = \sum_{i_1+i_2 +  i_3=r} \binom{r}{i_1, i_2, i_3}x^{i_1}_1x^{i_2}_2x^{i_3}_3  
$$
where 
\begin{equation}
\binom{r}{i_1, i_2, i_3} = \frac{(i_1+ i_2 + i_3)!}{i_1!i_2! i_3!}, \quad i_1+i_2+i_3=r
\end{equation}
is the tri-nomial coefficients, which  is symmetric for a permutation of $i_1, i_2, i_3$ and can be expressed in terms of binomial coefficients:
\begin{equation}
\binom{r}{i_1, i_2, i_3} =\binom{i_1+i_2}{i_1}\binom{i_1+i_2+i_3}{i_1+i_2}.
\end{equation} 
Therefore, for $r=n-s, p = 3, i_1 = s, i_2 = k-s, i_3 = n-k-s, i_1+i_2+i_3=n-s$, we can obtain
\begin{equation}\label{eq-tri-nomial-doeff}
\binom{n-s}{s, k-s, n-k-s} 
= \binom{k}{s}\binom{n-s}{k}. 
\end{equation}

When computing the value of binomial coefficients with computer programs, we should keep an eye on the overflow problem since the factorial $r!$ increases rapidly with the integer $r$. We can avoid such a risk by reformulating the binomial coefficient properly, i.e., for any $ i\in \mathbb{Z} = \set{0}\cup \mathbb{N}$, 
\begin{equation} \label{eq-psi-x-i}
\binom{\alpha}{i} = \left\{
\begin{array}{ll}
1, & i = 0; \\
\prod\limits^{i-1}_{t=0}\cfrac{\alpha-t}{i-t}, &  i\ge 1.
\end{array}
\right.
\end{equation}
The value of $\binom{\alpha}{i}$ can be computed with \Algr \ref{alg-binom} robustly and fastly.

\subsubsection{Kronecker Symbol}

The notation
\begin{equation}
\delta_{ij} = \left\{
\begin{array}{ll}
0, & i \neq j\\
1, & i = j
\end{array}
\right.
\end{equation}
is called the Kronecker symbol. Particularly, $\delta_{m0} = 1$ for $m = 0$ and $\delta_{m0}=0$ otherwise.

\subsubsection{Discrete Unit Step Function}
The discrete unit step function is defined by
\begin{equation}
u(m) =\left\{
\begin{array}{ll}
1, & m\in \mathbb{Z}^+=\set{0, 1, 2, 3, \cdots} \\
0, & m \in \mathbb{Z}^-=\set{-1, -2, -3, \cdots}
\end{array}
\right.
\end{equation}
with the counterpart of Heaviside function in the continuous case.

\subsubsection{Cardinality}

For a finite set $A=\set{x_1, \cdots, x_r}$, its cardinality is denoted by $\card{A}$, i.e.
\begin{equation}
\card{A} = \card{\set{x_1, \cdots, x_r}} = r.
\end{equation} 

\subsection{Zernike Circular Polynomials}  

\subsubsection{Definition of Zernike Circular Polynomials}

Generally, for $n\in\mathbb{Z}^+$ and $m\in \mathbb{Z}$ such that $\abs{m}\le n$ and $2\mid (n-m)$, the ZCP can be denoted by \cite{BornWolf-1999, Jasssen-2010, Shakibaei-2013}
\begin{equation}
\Zernpoly{n}{m}(\rho,\theta) 
= N^m_n \Radipoly{n}{m}(\rho)\Theta_m(\theta), \quad \rho \in [0, 1], \theta\in [0, 2\pi]
\end{equation}
where
\begin{equation} \label{eq-Z-Nnm}
N^m_n = \sqrt{\frac{2(n+1)}{1+\delta_{m0}}} =
\left\{
\begin{array}{ll}
\sqrt{2(n+1)}, & m \neq 0\\
\sqrt{n+1},    & m = 0
\end{array}
\right.
\end{equation}
is the coefficients for normalization,
\begin{equation} \label{eq-Z-Theta}
\Theta_m(\theta)=\left\{
\begin{array}{ll}
1,                   & m = 0, j\in \scrd{\mathbb{Z}}{odd}^+  \\                  
\cos(\abs{m}\theta), & m \neq 0, j\in \scrd{\mathbb{Z}}{even}^+  \\
\sin(\abs{m}\theta), & m \neq 0, j\in \scrd{\mathbb{Z}}{odd}^+  
\end{array}
\right.
\end{equation}
is the angular function with equivalent complex form $\me^{\mi \abs{m}\theta}$,  
\begin{equation} \label{eq-Z-Rnm}
\begin{split}
 \Radipoly{n}{m}(\rho)
 &= \frac{1}{(\frac{n-\abs{m}}{2})!\rho^{\abs{m}}} \left[\frac{\dif}{\dif (\rho^2)} \right]^{\frac{n-\abs{m}}{2}}
    \left[ (\rho^2)^{\frac{n+\abs{m}}{2}} (\rho^2 -1)^{\frac{n-\abs{m}}{2}} \right] \\
 &= \sum_{s=0}^{\frac{n-\abs{m}}{2}} (-1)^s \cdot
     \frac{(n-s)!}{s!\left(\frac{n+\abs{m}}{2} -s\right)!\left(\frac{n-\abs{m}}{2} -s\right)!} \cdot
       \rho^{n-2s} \\
 &=\sum_{s=0}^k (-1)^s \cdot \binom{n-s}{s,k-s,n-k-s} \cdot \rho^{n-2s}\\
 &=\sum_{s=0}^k c_s \rho^{p_s} = c_0\rho^n + c_1\rho^{n-2} + \cdots + c_k \rho^{n-2k}
\end{split}
\end{equation}
is the radial Zernike polynomials in which
\begin{equation} \label{eq-k}
k = \frac{1}{2}(n-\abs{m})
\end{equation}
and 
\begin{equation} \label{eq-k-cs}
\left\{
\begin{aligned}
c_s &= (-1)^s\binom{n-s}{s,k-s,n-k-s}, \quad 0 \le s \le k;\\ 
p_s &= n -2s, \quad  0 \le s \le k.
\end{aligned}
\right.
\end{equation} 
Substituting \eqref{eq-tri-nomial-doeff} into \eqref{eq-k-cs}, we obtain the following efficient expression for computing $c_s$
\begin{equation} \label{eq-cs}
c_s = (-1)^s \binom{k}{s}\binom{n-s}{k}, \quad 0\le s \le k.
\end{equation}

\subsubsection{Single and Double Indices for Zernike Circular Polynomials}

There are different schemes for representing the order of ZCP in theory analysis and applications. The first scheme is the \textit{Born-Wolf notation} (BWN), which uses the double indices $\mpair{n}{m}$ in the expression $\Zernpoly{n}{m}(\rho, \theta)$. This kind of notation is well known in the famous book by Born-Wolf \cite{BornWolf-1999}. The BWN is also named with OSA notation. 

The second scheme is characterized by a single index $j$. However, there are three choices for the single index:
\begin{itemize}
\item ANSI "Standard Zernike Polynomials". In this case $  \scrd{j}{_{ANSI}} = \frac{n(n+2)+m}{2}$.
\item ZEMAX "Standard Zernike Coefficients". This is introduced by Noll in studying atmospheric turbulence \cite{Noll-1976}, in which $j$ is used to represent the order instead of $n$ and $m$. In the software ZEMAX for optics design, the Noll's notation $j$ is taken.
\item ZEMAX "Zernike Fringe Polynomials" \footnote{It is also named polynomials of University of Arizona.}, which is introduced by James C. Wyant \cite{WyantZernike}.
\end{itemize} 
In this paper, we just concern the Noll's $Z_j$ and the Born-Wolf's $\Zernpoly{n}{m}$ due to their wide applications in optics design where the relation between $\mpair{n}{m}$ and $j$ is well known for optics engineers. In this sense, we have \cite{Noll-1976,Zemax-2008}
\begin{equation}
\Zern_j(\rho,\theta)= \Zernpoly{n}{m}(\rho, \theta), \quad j = j(n,m)\in \mathbb{N}
\end{equation}
such that
\begin{equation}
\braket{\Zern_j}{\Zern_{j'}} =\pi \delta_{jj'}.
\end{equation}

\subsubsection{Problem of the Inter-conversion of Single/Double Indices}

For the given single index $j$, we have \cite{OptikShopTest-2007, Mahajan-2013-vol-3}
\begin{equation}
n = n(j)= \mfloor{\sqrt{2j-1}+\frac{1}{2}} -1
\end{equation}
and
\begin{equation}
 m = m(j)
 =\left\{
 \begin{array}{ll}
 2\mfloor{\frac{2j+1 -n(n+1)}{4}}, &  n\in \scrd{\mathbb{Z}}{even}^+; \\
 2\mfloor{\frac{2j+1 -n(n+1)}{4}} -1, &  n\in \scrd{\mathbb{Z}}{odd}^+.
 \end{array}
 \right.
\end{equation}
However, there is a lack of simple expression to calculate $j$ when both $n$ and $m$ are given. Actually, what we can find is a description of $j$ in a range $ n(n+1)/2 +1\le j \le n(n+1)/2 + n + 1$ (as explained in \cite{OptikShopTest-2007, Mahajan-2013-vol-3}). Here both the lower and the upper bound for $j$ is not tight, so it can not be used to determine $j$ uniquely. Consequently, it is worth  exploring new results for the inter-conversion of $j$ and $\mpair{n}{m}$.

\subsection{Elements of a Table}

\subsubsection{General Description of a Table}

For a general table with abstract data set
$$
\mathscr{T} = \set{d^\alpha\in S_1\times S_2 \times \cdots \times S_{\scrd{n}{cols}}: \scrd{i}{begin} \le \alpha \le  \scrd{i}{end}}
$$
where $d^\alpha = (d^\alpha_1, \cdots, d^\alpha_\beta, \cdots, d^\alpha_{\scrd{n}{cols}}) $ is the $\alpha$-th record 
as shown in \Tab \ref{tab-general-form}, there are some fundamental  parameters for specifications: the caption of the table, the number of columns denoted  by $\scrd{n}{cols}$, the number of rows denoted by $\scrd{n}{rows}=\scrd{i}{end}-\scrd{i}{begin} + 1$, the list of attribute names with the size $\scrd{n}{cols}$, the elements of attribute data which have $\scrd{n}{rows}$ records and each record is an array of abstract data type (ADT)  with $\scrd{n}{cols}$ members.

\begin{table*}[h]
\centering
\caption{General form of caption for a table with $\scrd{n}{cols}$ attributes and $\scrd{n}{rows}=\scrd{i}{end}-\scrd{i}{begin}+1$ records}
\label{tab-general-form}
\begin{tabular}{llllll}
\hline
attribname-{\#}1 & attribname-{\#}2 & $\cdots$ & attribname-{\#}$\beta$ & $\cdots$  & attribname-{\#}$\scrd{n}{cols}$ \\
\hline
$d^{\scrd{i}{begin}}_1$  & $d^{\scrd{i}{begin}}_2$ & $\cdots$ & $d^{\scrd{i}{begin}}_\beta$  &   $\cdots$ &  $d^{\scrd{i}{begin}}_{\scrd{n}{cols}}$ \\
$d^{\scrd{i}{begin}+1}_1$  & $d^{\scrd{i}{begin}+1}_2$ & $\cdots$ & $d^{\scrd{i}{begin}+1}_\beta$  &   $\cdots$ &  $d^{\scrd{i}{begin}+1}_{\scrd{n}{cols}}$ \\
\quad $\vdots$ & \quad $\vdots$ &  $\cdots$  & \quad $\vdots$   & $\cdots$  & \quad$\vdots$ \\
$d^{\alpha}_1$  & $d^\alpha_2$ & $\cdots$ & $d^\alpha_\beta$  &   $\cdots$ &  $d^\alpha_{\scrd{n}{cols}}$ \\
\quad$\vdots$ & \quad$\vdots$ & $\cdots$  & \quad $\vdots$   & $\cdots$  & \quad $\vdots$ \\
$d^{{\scrd{i}{end}}}_1$  & $d^{{\scrd{i}{end}}}_2$ & $\cdots$ & $d^{{\scrd{i}{end}}}_\beta$  &   $\cdots$ &  $d^{{\scrd{i}{end}}}_{\scrd{n}{cols}}$ \\
\hline
\end{tabular}
\end{table*}

\subsubsection{Table of Mathematical Expressions for Zernike Circular Polynomials}

For our objective of automatically generating the long table 
\begin{equation}
\scrd{\mathscr{T}}{Zernike} 
= \set{d^j=[j, n, m, N^m_n, \Radipoly{n}{m}(\rho)\Theta_m(\theta)]: \scrd{j}{min}\le j \le \scrd{j}{max}},
\end{equation}
for the ZCP with algorithms and computer programs, 
all of the data terms in the table are strings specified with \LaTeX{} grammar. It is easy for us to set the following specifications:
\begin{itemize}
\item the number of columns is $\scrd{n}{cols} = 5$;
\item the list of attribute names are: 
\begin{itemize}
\item  attribname-{\#}1 --- $j$, the symbolic expression for the single index $j$ could be the string \lstinline|"$j$"|;  
\item  attribname-{\#}2 --- $n$, the symbolic expression for the  $n$-component of double indices $\mpair{n}{m}$ could be the string \lstinline|"$n$"|;
\item  attribname-{\#}3 --- $m$, the symbolic expression for the $m$-component of double indices $\mpair{n}{m}$ could be the string 
\lstinline|"$m$"|;
\item  attribname-{\#}4 --- $N^m_n$, the symbolic expression for the normalization coefficient could be the string \lstinline|"$N^m_n$"|;
\item  attribname-{\#}5 --- $\Radipoly{n}{m}(\rho)\Theta_m(\theta)$, the symbolic expression for the Zernike circular polynomial such that $\Zern_j(\rho,\theta) = \Radipoly{n}{m}(\rho)\Theta_m(\theta)$ could be 
the string \lstinline|"$\\Zern_j(\\rho,\\theta) = \\Radipoly{n}{m}(\\rho)\\Theta_m(\\theta)$"|; 
\end{itemize}
\item the caption of the table could be 
\begin{itemize}
\item caption --- Zernike Polynomials $\Zern_j(\rho,\theta)=N^m_n\Radipoly{n}{m}(\rho)\Theta_m(\theta)$.
\end{itemize}
\end{itemize}
As an illustration, for $1\le j \le 465$, our table (obtained by compiling the \LaTeX{} source file) will have the form like \Tab \ref{tab-Zernike-1-15}.

\begin{table*}[h]
\centering
\caption{Zernike Circular Polynomials $\Zern_j(\rho,\theta)=N^m_n\Radipoly{n}{m}(\rho)\Theta_m(\theta)$}
\label{tab-Zernike-1-15}
\begin{tabular}{ccrcp{0.55\textwidth}}
\hline
$j$ & $n$ & $m$ & $N^m_n$ &  $\Radipoly{n}{m}(\rho)\Theta_m(\theta)$\\
\hline
 $1$  & $0$  & $0$  &$\sqrt{1}$  &$1 $\\
 $2$  & $1$  & $-1$  &$\sqrt{4}$  &$\rho \cos(\theta)$\\
 $3$  & $1$  & $1$  &$\sqrt{4}$  &$\rho \sin(\theta)$\\
 $4$  & $2$  & $0$  &$\sqrt{3}$  &$2\rho^{2} -1 $\\
 $5$  & $2$  & $-2$  &$\sqrt{6}$  &$\rho^{2} \sin(2\theta)$\\
 $6$  & $2$  & $2$  &$\sqrt{6}$  &$\rho^{2} \cos(2\theta)$\\
 $7$  & $3$  & $-1$  &$\sqrt{8}$  &$(3\rho^{3} -2\rho )\sin(\theta)$\\
 $8$  & $3$  & $1$  &$\sqrt{8}$  &$(3\rho^{3} -2\rho )\cos(\theta)$\\
 $9$  & $3$  & $-3$  &$\sqrt{8}$  &$\rho^{3} \sin(3\theta)$\\
 $10$  & $3$  & $3$  &$\sqrt{8}$  &$\rho^{3} \cos(3\theta)$\\
 $11$  & $4$  & $0$  &$\sqrt{5}$  &$6\rho^{4} -6\rho^{2} +1 $\\
 $12$  & $4$  & $-2$  &$\sqrt{10}$  &$(4\rho^{4} -3\rho^{2} )\cos(2\theta)$\\
 $13$  & $4$  & $2$  &$\sqrt{10}$  &$(4\rho^{4} -3\rho^{2} )\sin(2\theta)$\\
 $14$  & $4$  & $-4$  &$\sqrt{10}$  &$\rho^{4} \cos(4\theta)$\\
 $15$  & $4$  & $4$  &$\sqrt{10}$  &$\rho^{4} \sin(4\theta)$\\
 $\vdots$ & $\vdots$ & $\vdots$ & $\vdots$ & $\vdots$ \\
$46$  & $9$  & $-1$  &$\sqrt{20}$  &$(126\rho^{9} -280\rho^{7} +210\rho^{5} -60\rho^{3} +5\rho )\cos(\theta)$\\
 $47$  & $9$  & $1$  &$\sqrt{20}$  &$(126\rho^{9} -280\rho^{7} +210\rho^{5} -60\rho^{3} +5\rho )\sin(\theta)$\\
 $\vdots$ & $\vdots$ & $\vdots$ & $\vdots$ & $\vdots$ \\ 
  $464$  & $29$  & $-29$  &$\sqrt{60}$  &$\rho^{29} \cos(29\theta)$\\
 $465$  & $29$  & $29$  &$\sqrt{60}$  &$\rho^{29} \sin(29\theta)$\\
\hline
\end{tabular}
\end{table*}

\section{Inter-conversion of Single/Double Indices of Zernike Circular Polynomials} \label{sec-newprinciple}

The inter-conversion of the double indices $\mpair{n}{m}$ and single index $j$ is significant for generating the table of ZCP  and looking up the table for the polynomials of interests automatically. In this section, we will establish new formula and theorems to specify the implementation of the inter-conversion for Noll's single index $j$ and Born-Wolf's double indices $\mpair{n}{m}$.

\subsection{Conversion of Double/Single Indices}

The Noll's index $j$  and the Born-Wolf's indices $\mpair{n}{m}$ can be converted to each other and the conversions can be uniquely determined.  

\begin{thm}
The Noll's index $j$ can be computed with Born-Wolf's indices $\mpair{n}{m}$ by
\begin{equation} \label{eq-nm2j}
j(n,m) = \frac{n(n+1)}{2} + \abs{m} + u(m)
\end{equation}
for $\abs{m} \le n$ and $2 \mid (n-m)$.
\end{thm}

\noindent \textbf{Proof}: Let 
\begin{equation}
\begin{aligned}
V_n&= \langle v_1, \cdots, v_r, \cdots, v_{n+1}\ \rangle \\
 &= \left\{
            \begin{array}{ll}
              \langle 0,-2, 2, -4, 4, \cdots, -n, n \rangle, &  n \in \scrd{\mathbb{Z}}{even}^+ \\
              \langle -1, 1, -3, 3, \cdots, -n, n \rangle,   &  n \in \scrd{\mathbb{Z}}{odd}^+
            \end{array}
          \right.
\end{aligned}
\end{equation}
be the sequence of possible integers $m$ such that
\begin{equation} \label{eq-r2m}
m = v_r = \left\{
\begin{array}{ll}
r-1, & \mathrm{for}~ 2 \nmid r~\mathrm{and}~ 2\mid n; \\
-r,  & \mathrm{for}~ 2 \mid r~\mathrm{and}~ 2\mid n ; \\
-r, & \mathrm{for}~ 2\nmid  r~\mathrm{and}~ 2\nmid n; \\
r-1,  & \mathrm{for}~ 2\mid r~\mathrm{and}~ 2\nmid n.
\end{array}
\right. 
\end{equation}
In other words, the integer 
\begin{equation} \label{eq-rm}
r = \abs{m} + u(m) =\left\{
\begin{array}{ll}
m + 1, & m \ge 0\\
-m, & m <0
\end{array}
\right.
\end{equation}
is the subscript $r$ or the position in the sequence $V_n$ for $m = v_r\in V_n$. It is easy to find that the cardinality of $V_n$ is 
\begin{equation}
\card{V_n} = n+1, \quad n\in \mathbb{Z}^+
\end{equation}
Therefore, for $i\in \set{0, 1, \cdots, n-1}$, the total number of Zernike polynomial $\Zernpoly{i}{m}(\rho, \theta)$ is
$$
\sum^{n-1}_{i=0}\card{V_{i}} = \sum^{n-1}_{i=0} (i+1) =  1+ 2 + \cdots + n = \frac{n(n+1)}{2}.
$$ 
Thus for the given $n$ and $m$, let $r$ be position of $m\in V_{n\mod 2}$, then the corresponding index $j$ for the 
$\Zern_j(\rho,\theta) = \Zernpoly{n}{m}(\rho, \theta)$ must be the sum of base position $n(n+1)/2$ and relative position $r$, i.e., 
\begin{equation} \label{eq-def-r}
j = \frac{n(n+1)}{2} + r
\end{equation}
This implies that \eqref{eq-nm2j} holds since we have \eqref{eq-rm}. Q.E.D.

\subsection{Conversion of Single/Double Indices}

As mentioned above, there is a lack of simple formulae for converting Noll's single index $j$ to Born-Wolf's double indices $\mpair{n}{m}$. The following theorem remedies the defects.  

\begin{thm}
Given the Noll index $j$, the Born-Wolf's indices $\mpair{n}{m}$ can be computed with the following
expressions:
\begin{align}
n(j) &=  \mceil{\frac{-3+\sqrt{8j+1}}{2}},   \label{eq-j2n}\\
r(j) &= j - \frac{n(j)[n(j)+1]}{2},  \label{eq-j2r}\\
m(j) &= v_{r(j)} = \left\{
\begin{array}{ll}
r(j)-1, & \mathrm{for}~ 2\nmid r(j)~\mathrm{and}~ 2\mid n(j); \\
-r(j),  & \mathrm{for}~2\mid r(j)~\mathrm{and}~ 2\mid n(j); \\
-r(j), & \mathrm{for}~2\nmid r(j)~\mathrm{and}~2 \nmid n(j); \\
r(j)-1,  & \mathrm{for}~2\mid r(j)~\mathrm{and}~2\nmid n(j).
\end{array}
\right.   
\label{eq-j2m}
\end{align}
\end{thm}

\noindent{\textbf{Proof}}: By (\ref{eq-nm2j}), we have
\begin{equation}
\frac{n(n+1)}{2} < j \le \frac{n(n+1)}{2} + n+1.
\end{equation}
Consequently,
\begin{equation}
 \frac{-3 + \sqrt{8j+1}}{2} \le n < \frac{-1 + \sqrt{8j+1}}{2}
\end{equation}
This implies \eqref{eq-j2n} by the definition of floor of real number. \eqref{eq-j2r} is obvious by \eqref{eq-def-r}.
With the help of \eqref{eq-r2m} we immediately have \eqref{eq-j2m}. Q.E.D.

\section{Generating Table of Zernike Circular Polynomials} 
\label{sec-gen-table}

\subsection{Numeric Computation for Generating Zernike Circular Polynomials}

The purpose of numeric computation for generating ZCP is to determine the key parameters involved, which includes single/double indices $j, n, m$, binomial coefficients $\binom{\alpha}{i}$ and the powers $p_i$ in radial polynomial function $\Radipoly{n}{m}(\rho)$. 

\subsubsection{Calcualte the Parameter $k$ for the Radial Zernike Polynomial}

The parameter $k$ in \eqref{eq-k} determines the number of terms of the radial Zernike polynomial $\Radipoly{n}{m}(\rho)$. The $k$ can be directly obtained by the double indices $\mpair{n}{m}$, see \Algr \ref{alg-nm2k}.

\begin{breakablealgorithm}
\caption{Calculate the parameter $k$ for the number of terms in the radial Zernike polynomial $\Radipoly{n}{m}(\rho)$}
\label{alg-nm2k} 
\begin{algorithmic}[1]
\Require Noll's double indices $\mpair{n}{m}$
\Ensure  the parameter $k$ for the radial polynomial $\Radipoly{n}{m}(\rho)$
\Function{CvtNM2K}{$\mpair{n}{m}$}
\State $k \gets (n-\abs{m})/2$;
\State \Return $k$;
\EndFunction 
\end{algorithmic}
\end{breakablealgorithm}

\subsubsection{Interconversion of Single/Double Indices}

Given the BWN pair $\mpair{n}{m}$, the Noll's single index $j$ can be determined by \eqref{eq-nm2j}, please see the procedure \ProcName{CvtNM2J} in \Algr \ref{alg-nm2j} for the details. On the other hand, suppose the single index $j$ is known, we can compute the double indices $\mpair{n}{m}$ according to \eqref{eq-j2n}, \eqref{eq-j2r} and \eqref{eq-j2m}. The procedure \ProcName{CvtJ2NM} in \Algr \ref{alg-j2nm} illustrates the steps and details completely.

\begin{breakablealgorithm}
\caption{Convert the BWN pair $\mpair{n}{m}$ to the Noll's single index $j$}
\label{alg-nm2j}
\begin{algorithmic}[1]
\Require Double indices $\mpair{n}{m}$ of BWN where $n\in \mathbb{Z}^+$ and $m\in \mathbb{Z}$ such that $-n \le m \le n$ and $2\mid (n-m)$.
\Ensure Single index $j\in \mathbb{N}$ of Noll notation.
\Function{CvtNM2J}{$\mpair{n}{m}$}
\State Check the input $n$ and $m$: $n\ge 0, -n\le m \le n, 2\mid (n-m)$.
\If{($m \ge 0$)}
   \State $j \gets n(n+1)/2 + m + 1$;
\Else
   \State   $j \gets n(n+1)/2 -m$;
\EndIf
\State \Return $j$; 
\EndFunction
\end{algorithmic}
\end{breakablealgorithm}

\begin{breakablealgorithm}
\caption{Convert the single index $j$ of Noll notation to double indices $\mpair{n}{m}$ of BWN}
\label{alg-j2nm}
\begin{algorithmic}[1]
\Require Single index $j$ of Noll notation.
\Ensure  BWN pair $\mpair{n}{m}$ where $n\in \mathbb{Z}^+$ and $m\in \mathbb{Z}$ such that $\abs{m} \le n$ and $2\mid (n-m)$.
\Function{CvtJ2NM}{$j$}
\State Check the value of the integer $j$: $j\ge 1$
\State $n \gets  \mceil{(-3+\sqrt{8j+1})/2}$;
\State $r \gets j - n(n+1)/2$;
\If{($2\mid n$)}
  \If{($2\mid r$)}
      \State $m \gets -r$;  \quad // For $2\mid n$ and $2\mid r$, $m = -r$.
   \Else
      \State $m \gets r-1$; // For $2\mid n$ and $2\nmid r$, $m = r-1$.
   \EndIf
\Else
  \If{($2\mid r$)}
    \State $m\gets r - 1$; // For $2\nmid n$ and $2\mid r$, $m = r-1$.
  \Else
    \State $m \gets -r$;    \quad // For $2\nmid n$  and $2\nmid r$, $m = -r$.
  \EndIf
\EndIf
\State \Return $\mpair{n}{m}$;
\EndFunction
\end{algorithmic}
\end{breakablealgorithm}

\subsubsection{Computation of Binomial Coefficients}

The binomial coefficients $\binom{\alpha}{i}$ is important for computing the coefficients $c_s$ in \eqref{eq-cs}. The robust and fast procedure \ProcName{CalcBinomCoef} for computing the $\binom{\alpha}{i}$ is given in \Algr \ref{alg-binom}.

\begin{breakablealgorithm}
\caption{Calcluating the binomial coefficient $\displaystyle \binom{\alpha}{i}$} \label{alg-binom}
\begin{algorithmic}[1]
\Require Real number $\alpha\in \mathbb{R}$ and non-negative integer $i\in \mathbb{Z}^+$;
\Ensure The real number  $\displaystyle \binom{\alpha}{i}$.
\Function{CalcBinomCoef}{$\alpha, i$}
\State Check the value of integer $i$: $i\ge 0$
\State $\cpvar{prod}\gets 1$;
\If{($i\ge 1$)}
\For{$t\in\seq{0, 1, \cdots, i-1}$}
\State $\cpvar{prod}\gets \cpvar{prod}\cdot (\alpha-t)/(i-t)$;
\EndFor
\EndIf
\State \Return \cpvar{prod};
\EndFunction
\end{algorithmic}
\end{breakablealgorithm}

\subsubsection{Computation of Coefficients and Power Indices of Radial Zernike Polynomials}

The coefficients $\set{c_s: 0\le s\le k}$ and power indices $\set{p_s: 0\le s\le k}$ in the radial Zernike polynomials can be computed by the procedure \ProcName{CalcRadialPolyPowerCoef} in \Algr \ref{alg-radipoly-coef-power} according to \eqref{eq-k-cs} and \eqref{eq-cs}. 

\begin{breakablealgorithm}
\caption{Calculate the sequence of coefficients $\cpvar{c}= \seq{c_0, c_1, \cdots, c_k}$ and sequence of power indices $\cpvar{p} = \seq{p_0, p_1, \cdots, p_k}$ for the radial polynomial function $\Radipoly{n}{m}(\rho)$}  
\label{alg-radipoly-coef-power}
\begin{algorithmic}[1]
\Require BWN pair $\mpair{n}{m}$ where $n\in \mathbb{Z}^+$ and $m\in \mathbb{Z}$ such that $\abs{m} \le n$ and $n-m$ is even.
\Ensure The sequence of coefficients $\cpvar{c}=\seq{c_0, c_1, \cdots, c_k}$ and the sequence of powers $\cpvar{p} = \seq{p_0, p_1, \cdots, p_k}$ of radial polynomial function $\Radipoly{n}{m}(\rho)$ where $k = (n-\abs{m})/2$.
\Function{CalcRadialPolyPowerCoef}{$\mpair{n}{m}$}
\State $k\gets \ProcName{CvtNM2K}(\mpair{n}{m})$;
\State $\cpvar{sign} \gets 1$;
\For{$s\in \seq{0, 1, \cdots, k}$}
\State $c_s \gets \cpvar{sign}\cdot \ProcName{CalcBinomCoef}(k,s)\cdot\ProcName{CalcBinomCoef}(n-s, k)$;
\State $p_s \gets n-2s$;
\State $\cpvar{sign} \gets -\cpvar{sign}$;
\EndFor
\State \Return  $\mpair{\cpvar{c}}{\cpvar{p}}$;
\EndFunction
\end{algorithmic}
\end{breakablealgorithm}

\subsection{Symbolic Computation for Generating Zernike Circular Polynomials}

\subsubsection{Generating Symbolic Expression for a Polynomial}

\begin{figure}[h]
	\centering
	\begin{forest}
		[\fbox{\ProcName{GenSexprPolynom}}
		[\fbox{\ProcName{GenSexpreGeneralTerm}}
		   [\fbox{\ProcName{GenSexprCatSign}}]
		   [\fbox{\ProcName{GenSexprUsgnGeneralTerm}}]		
		]
		] 
	\end{forest}
	\caption{Procedures involved in generating symbolic expression for a polynomial.}
	\label{fig-Sexpr-polynom-procedurs}
\end{figure}
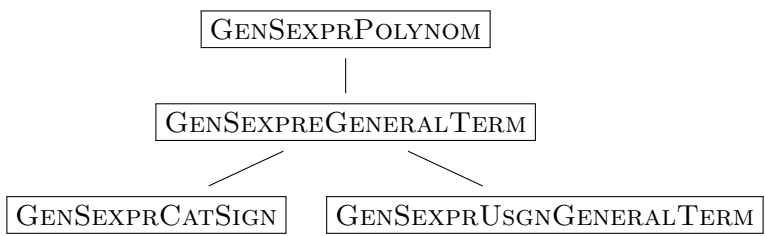

Formally, there are two steps for generating the symbolic expression for a polynomial $g(x) = \sum\limits_{s=0}^k c_s x^{p_s}$:
\begin{itemize}
\item generate the symbolic expression for the general term $c_sx^{p_s}$
\begin{itemize}
\item generate the unsigned general term $ \abs{c_s}x^{p_s}$;
\item determine the sign of $c_s$ since we may get \lstinline|"+"|, \lstinline|"-"| or \lstinline|"0"|; 
\end{itemize}
\item generate all of the $k+1$ terms one by one with an iterative operation via loop construction in high level computer programming language.
\end{itemize}
\Fig \ref{fig-Sexpr-polynom-procedurs} illustrates these steps with the help of nesting procedures for sub-tasks.

%\begin{figure}[h]
%	\centering
%    \includegraphics[width=0.5\textwidth]{Fig-2.png} 
%	\caption{Procedures involved in generating symbolic expression for a polynomial.}
%	\label{fig-Sexpr-polynom-procedurs}
%\end{figure}

When the sequence of coefficients  $\cpvar{c}=\seq{c_0, c_1, \cdots, c_k}$ and the sequence of power indices $\cpvar{p}=\seq{p_0, p_1, \cdots, p_k}$ of the polynomial $f(x) = c_0 x^{p_0} + \cdots c_sx^{p_s} + \cdots + c_k x^{p_k}$ are given, the symbolic expression 
of $g(x)$ can be generated with an iterative process:
\begin{itemize}
\item determining the generating method for the symbolic expression of general term $c_sx^{p_s}$;
\item generating all of the terms iteratively for $s=0, 1, 2, \cdots, k$.
\end{itemize}
However, there are three issues to be settled: 
\begin{itemize}
\item specifying the formal variable $x$ for the polynomial since it may be $x$, $t$, $\rho$ or other possible symbol; 
\item specifying the operator for representing the power since $x^r$ could be implemented by \verb|x^r| in 
\LaTeX{} and MATLAB/Octave, or by \verb|x**r| for Python, and so on;
\item specifying the special cases for the general term:
      \begin{itemize}
      \item if $p_s = 1$, then $c_sx^1$ should be replaced by $c_sx$;
      \item if $p_s = 0$, then $c_sx^0$ should be replaced by $c_s$;
      \item if $c_s < 0$, then the string $+c_sx^{p_s}$ should be replaced by $-\abs{c_s}x^{p_s}$;
      \item if $c_s = 0$, then the term $c_sx^{p_s}$ should disappear and be ignored.  
      \end{itemize}
\end{itemize}
For the mathematical expression of polynomials, we should consider the coefficients and their signs carefully. The procedure \ProcName{GenSexprCatSign} in \Algr \ref{alg-sign-cat} is used to determine the symbol "$+$" or "$-$" for concatenation. The steps for generating a general polynomial are built on the following operations:
\begin{itemize}
\item determining the symbol "$+$" or "$-$" for the $i$-th term $c_ix^{p_i}$ according to the procedure \ProcName{GenSexprCatSign};
\item generating symbolic expressions for the term $\abs{c_i}x^{p_i}$ without the sign of $c_i$ via the procedure \ProcName{GenSexprUsgnGeneralTerm} in \Algr \ref{alg-abs-term};
\item generating the symbolic expression for the general term $c_ix^{p_i}$ in the polynomial according to the procedure \ProcName{GenSexprGeneralTerm} in \Algr \ref{alg-general-term};
\item generating the polynomial $\displaystyle f(x) = \sum_i c_ix^{p_i}$ with the procedure \ProcName{GenSexprPolynom} in \Algr \ref{alg-Sexpr-polynom}.
\end{itemize} 
It should be noted that if $c_i= 0$, the term $c_ix^{p_i}$ will be ignored for generating symbolic expression. 

\begin{breakablealgorithm}
\caption{Generate the symbol "$+$" or "$-$" for concatenation}
\label{alg-sign-cat}
\begin{algorithmic}[1]
\Require Argument \cpvar{fp} for text file, Coefficient $c_i$ and position label $i$
\Ensure The sign of $c_i$
\Function{GenSexprCatSign}{\cpvar{fp}, $c_i$, $i$}
\If{($c_i < 0$)}
\State \ProcName{Fprintf}(\cpvar{fp}, "$-$");
\Else \quad // for $c_i \ge 0$
   \If{($i \neq 0 $)}  
   \State \ProcName{Fprintf}(\cpvar{fp}, "+"); 
   \EndIf
\EndIf
\EndFunction
\end{algorithmic}
\end{breakablealgorithm}

\begin{breakablealgorithm}
\caption{Generate symbolic expression for the term $\abs{c_i}x^{p_i}$ without the sign of $c$} \label{alg-abs-term}
\begin{algorithmic}[1]
\Require Argument $c_i$ for the $i$-th coefficient , argument $p_i$ for the power index;
\Ensure The string of characters for $\abs{c_i}x^{p_i}$; 
\Function{GenSexprUsgnGeneralTerm}{\cpvar{var}, \cpvar{op}, $c_i$, $p_i$}
\State $\cpvar{ucoef} \gets \abs{c_i}$;
\Switch{$p_i$}
\Case{$1$} 
   \If{($\cpvar{ucoef}\neq 1$)}
      \State \ProcName{Fprintf}(\cpvar{fp}, \lstinline|"%d%s",ucoef,var)|; 
   \Else 
       \State \ProcName{Fprintf}(\cpvar{fp}, \lstinline|"%s",var)|; 
   \EndIf
\EndCase
\Case{$0$}
   \State \ProcName{Fprintf}(\cpvar{fp}, \lstinline|"%d",ucoef)|;
\EndCase
\Default 
   \If{($\cpvar{ucoef}\neq 1$)}
      \State \ProcName{Fprintf}(\cpvar{fp}, \lstinline|"%d%s%s{%d}"|, \cpvar{ucoef}, \cpvar{var}, \cpvar{op}, $p_i$);
   \Else 
       \State \ProcName{Fprintf}(\cpvar{fp}, \lstinline|"%s%s{%d}"|, \cpvar{var}, \cpvar{op}, $p_i$);
   \EndIf
\EndDefault
\EndSwitch
\EndFunction
\end{algorithmic}
\end{breakablealgorithm}

\begin{breakablealgorithm}
\caption{Generate the symbolic expression for the general term $c_ix^{p_i}$ in the polynomial} 
\label{alg-general-term}
\begin{algorithmic}[1]
\Require Argument \cpvar{fp} for text file, integer $i$, string argument \cpvar{var} for $x$, string argument \cpvar{op}, argument $c_i$ for the $i$-th coefficient, integer argument for $p_i$ for the power index of single term
\Ensure Symbolic expression of $c_ix^{p_i}$ stored in the text file accessed by \cpvar{fp}.
\Function{GenSexprGeneralTerm}{\cpvar{fp}, $i$, \cpvar{var}, \cpvar{op}, $c_i$, $p_i$}
\State \ProcName{GenSexprCatSign}(\cpvar{fp}, $c_i$, $i$);
\State \ProcName{GenSexprUsgnGeneralTerm}(\cpvar{fp}, \cpvar{var}, \cpvar{op}, $c_i$, $p_i$);
\EndFunction
\end{algorithmic}
\end{breakablealgorithm}

\begin{breakablealgorithm}
\caption{Generate Symbolic Expression for a polynomial $\displaystyle f(x) = \sum_{s=0}^{k}c_sx^{p_s}$} \label{alg-Sexpr-polynom}
\begin{algorithmic}[1]
\Require Argument \cpvar{fp} for text file, symbolic argument \cpvar{var} for $x$, symbolic argument \cpvar{op} for the power operator, sequence of coefficients $\cpvar{c}=\seq{c_0, c_1, \cdots, c_k}$, sequence of power indices $\cpvar{p}=\seq{p_0, p_1, \cdots, p_k}$, integer \cpvar{size} for the counting number of single terms in $f(x)$.
\Ensure Symbolic expression for the polynomial $f(x)$ stored in the text file accessed by \cpvar{fp}.
\Function{GenSexprPolynom}{\cpvar{fp}, \cpvar{var}, \cpvar{op}, \cpvar{c}, \cpvar{p}, \cpvar{size}}
\For{$s\in \seq{0,1,\cdots, \cpvar{size}-1}$}
 \If{($c_s = 0$)}  // if $c_s$ is equal to 0
    \State \textbf{continue};  // just ignore the term $c_sx^{p_s}$ since it is zero, increase $s$ by 1
 \EndIf
 \State \ProcName{GenSexprGeneralTerm}(\cpvar{fp}, $s$, \cpvar{var}, \cpvar{op}, $c_s$, $p_s$);
 \State \ProcName{Fprintf}(\cpvar{fp}, \lstinline|"  ")|;   // is is optional , print empty space for better visualization. 
\EndFor
\EndFunction
\end{algorithmic}
\end{breakablealgorithm}

\subsubsection{Generating Symbolic Expressions for Radial Zernike Polynomials}

Once the algorithm for generating a general polynomial is designed, we can use it to generate the Zernike radial polynomials $\Radipoly{n}{m}(\rho)$ where the coefficients and power indices should be computed properly. The procedure \ProcName{GenSexprRadialPolynom} in \Algr \ref{alg-gen-radipoly}
is designed for this purpose.

\begin{breakablealgorithm}
\caption{Generating symbolic expression for the Zernike radial polynomial $\Radipoly{n}{m}(\rho)$} 
\label{alg-gen-radipoly}
\begin{algorithmic}[1]
\Require Argument \cpvar{fp} for text file, double indices $\mpair{n}{m}$, string \cpvar{var} for $\rho$, string \cpvar{op} for \lstinline|"^"|.
\Ensure Print symbolic expression for the Zernike Radial polynomial 
$\displaystyle \Radipoly{n}{m}(\rho) = \sum^{k}_{s=0} c_s\rho^{p_s}$.  
\Function{GenSexprRadialPolynom}{\cpvar{fp}, $\mpair{n}{m}$, \cpvar{var}, \cpvar{op}}
\State $\cpvar{size}\gets 1 + \ProcName{CvtNM2K}(\mpair{n}{m})$;\quad // \cpvar{size} = 1 + $k$;
\State $\mpair{\cpvar{c}}{\cpvar{p}}\gets \ProcName{CalcRadiPolynomPowerCoef}(\mpair{n}{m})$; \quad // \Algr \ref{alg-radipoly-coef-power}
\State \ProcName{GenSexprPolynom}(\cpvar{fp}, \cpvar{var}, \cpvar{op}, \cpvar{c}, \cpvar{p}, \cpvar{size});\quad // \Algr \ref{alg-Sexpr-polynom}
\EndFunction
\end{algorithmic}
\end{breakablealgorithm}

\subsubsection{Generating Symbolic Expressions for Angular Function}

The symbolic computation for the angular function $\Theta_m(\theta)$ is straight ford according to \eqref{eq-Z-Theta}, please see the procedure
\ProcName{GenSexprAnguFun} in \Algr \ref{alg-angular-function}.

\begin{breakablealgorithm}
\caption{Generate Symbolic Expression for Angular Function $\Theta_m(\theta)$} 
\label{alg-angular-function}
\begin{algorithmic}[1]
\Require Argument \cpvar{fp} for text file, double indices $\mpair{n}{m}$
\Ensure  Symbolic expression for angular function $\theta_m(\theta)$
\Function{GenSexprAnguFun}{\cpvar{fp}, $\mpair{n}{m}$}
\State $j \gets \ProcName{CvtNM2J}(\mpair{n}{m})$;
\State $m\gets \abs{m}$;
\If{($j\in \scrd{\mathbb{Z}}{even}$)}
   \Switch{$m$}
   		\Case{0}
   		    \If{($j = 0$)}
   		    \State   \ProcName{Fprintf}(\cpvar{fp}, \lstinline|"1"|);
   		    \EndIf
   		\EndCase
   		\Case{1}
   		    \State \ProcName{Fprintf}(\cpvar{fp}, \lstinline|"\\cos(\\theta)"|); \quad // print $\cos(\theta)$
   		\EndCase
   		\Default
   		    \State \ProcName{Fprintf}(\cpvar{fp}, \lstinline|"\\cos(%d\\theta)"|, $m$); \quad // print $\cos(m\theta)$
   		\EndDefault
   \EndSwitch
\Else
   \Switch{$m$}
   		\Case{0}
          \State    // do nothing
   		\EndCase
   		\Case{1}
   		    \State \ProcName{Fprintf}(\cpvar{fp}, \lstinline|"\\sin(\\theta)"|); \quad // print $\sin(\theta)$
   		\EndCase
   		\Default
   		    \State \ProcName{Fprintf}(\cpvar{fp}, \lstinline|"\\sin(%d\\theta)"|, $m$); \quad // print $\sin(m\theta)$
   		\EndDefault
   \EndSwitch
\EndIf
\EndFunction
\end{algorithmic}
\end{breakablealgorithm}

\subsubsection{Generating Symbolic Expressions for Normalization Coefficient}

The symbolic computation of the normalization coefficient $N^m_n$ can be based on \eqref{eq-Z-Nnm} and a simple if-else statement is enough, please see the procedure \ProcName{GenSexprRadiNormCoef} in \Algr \ref{alg-comp-Nnm}.

\begin{breakablealgorithm}
\caption{Generating symbolic expression for the normalization coefficient $N_n^m$}
\label{alg-comp-Nnm}
\begin{algorithmic}[1]
\Require Argument \cpvar{fp} for text file, double indices $\mpair{n}{m}$
\Ensure Print the symbolic expression normalization of coefficient $N^m_n=\sqrt{2(n+1)/(1+\delta_{0m})}$
\Function{GenSexprRadiNormCoef}{\cpvar{fp}, $\mpair{n}{m}$}
\If{$m=0$}
    \State \ProcName{Fprintf}(\cpvar{fp}, \lstinline|"\\sqrt{%d}"|, $n+1$); \quad // print $\sqrt{n+1}$
\Else
    \State \ProcName{Fprintf}(\cpvar{fp}, \lstinline|"\\sqrt{%d}"|, $2*(n+1)$); \quad // print $\sqrt{2(n+1)}$
\EndIf
\EndFunction
\end{algorithmic}
\end{breakablealgorithm}

\subsubsection{Generating Non-Standard (Unnormalized) Zernike Circular Polynomial}
For any single index $j\in \mathbb{Z}^+$ or the corresponding double indices $\mpair{n}{m}$, the ZCP $\Zern_j(\rho,\theta) = \Zern_n^m(\rho,\theta)=N^m_n \Radipoly{n}{m}(\rho)\Theta_m(\theta)$ consists of three parts:
the normalization coefficient $N^m_n$ determined by \eqref{eq-Z-Nnm},
the radial polynomial $\Radipoly{n}{m}(\rho)$ specified by \eqref{eq-Z-Rnm}, and the angular function $\Theta_m(\theta)$ given by \eqref{eq-Z-Theta}. The procedure \ProcName{GenSexprZernikePolynomNM} in \Algr 
\ref{alg-Gen-unnormalized-Znm} generates the un-normalized ZCP, which separates the normalization coefficient $N^m_n$ clearly.

\begin{breakablealgorithm}
\caption{Generate the \LaTeX{} code for the non-standard (un-normalized) Zernike circular function $\hat{Z}_j(\rho,\theta)=\Radipoly{n}{m}(\rho)\Theta_m(\theta)$} \label{alg-Gen-unnormalized-Znm}
\begin{algorithmic}[1]
\Require Argument \cpvar{fp} for text file, double indices $\mpair{n}{m}$
\Ensure Symbolic expression for $\Radipoly{n}{m}(\rho)\Theta_m(\theta)$ denoted by the grammar of \LaTeX{}
\Function{GenSexprZernikePolynomNM}{\cpvar{fp}, $\mpair{n}{m}$}
\State $\cpvar{size} \gets 1 + \ProcName{CvtNM2K}(\mpair{n}{m})$;
\State $\cpvar{var} \gets$ \lstinline|"\\rho"|; \quad // for the symbol $\rho$ in \LaTeX{}
\State $\cpvar{op} \gets$ \lstinline|"^"|; \quad // for the power operator in \LaTeX{}
\If{($m=0$ or $\cpvar{size}=1$)}
\State \ProcName{GenSexprRadiPolynom}(\cpvar{fp}, $\mpair{n}{m}$, \cpvar{var}, \cpvar{op}); \quad // print $\Radipoly{n}{m}(\rho)$
\Else
\State \ProcName{Fprintf}(\cpvar{fp},\lstinline|"("|); \quad // print left bracket
\State \ProcName{GenSexprRadiPolynom}(\cpvar{fp}, $\mpair{n}{m}$, \cpvar{var}, \cpvar{op}); 
\State \ProcName{Fprintf}(\cpvar{fp},\lstinline|")"|); \quad // print right braket
\EndIf
\State \ProcName{GenSexprAnguFun}(\cpvar{fp}, $\mpair{n}{m}$); \quad // print $\Theta_m(\theta)$
\EndFunction
\end{algorithmic}
\end{breakablealgorithm}

\subsection{Meta-Programming and \LaTeX{} Programming}

For the purpose of generating mathematical formula for ZCP, we can use meta-programming and \LaTeX{} programming. 
Essentially, the key idea of meta-programming is generating destination code with algorithms and computer programs implemented by source code. There are two fundamental steps to do so: 
\begin{itemize}
\item generating \LaTeX{} code (destination code) for creating symbolic expressions for Zernike circular polynomials with some high level programming languages such as C/C++, Octave/MATLAB, Python, Java and so on (source code);  
\item compiling the \LaTeX{} code to generate the symbolic expressions and output a file with the format *.dvi or *.pdf. 
\end{itemize}
In this work, we use the C programming language to generate the \LaTeX{} source code for representing the symbolic expressions for ZCP. Our emphasis is put on the algorithms instead of concrete C code since the algorithms can be implemented with various programming languages.

\subsubsection{Generating \LaTeX{} Code for a Line of a Table}

The procedure \ProcName{GenLaTeXTableLine} is used to generate a line $d^i$ of a long table 
$$\mathscr{T}=\set{d^i\in S_1\times S_2 \times \cdots \times S_{\scrd{n}{cols}}: \scrd{i}{begin}\le i\in \scrd{i}{end}}.$$
However, it depends on the concrete problem and we have to give details in the program. In object-oriented programming, it is a good idea to implement it with virtual function. For the purpose of generating ZCP, we can set $i\gets j$ (Noll's single index $j$), $\scrd{n}{cols} \gets 5$, 
$\scrd{i}{begin}\gets \scrd{j}{min}$ and $\scrd{i}{end}\gets \scrd{j}{max}$. \Algr \ref{alg-gen-table-line} gives the method of generating the $\Zern_j(\rho,\theta)$. 

\begin{breakablealgorithm}
\caption{Generate \LaTeX{} code for the line of a table of Zernike Circular Polynomials}
\label{alg-gen-table-line}
\begin{algorithmic}[1]
\Require Argument $\cpvar{fp}$ for text file, single index $j\in \mathbb{N}$
\Ensure  \LaTeX{} code for the $j$-th line  $d^j$, viz.,
$$d^j=[j,n,m,N^m_n,\Radipoly{n}{m}(\rho)\Theta_m(\theta)]\in \scrd{\mathscr{T}}{Zernike}$$
\Function{GenLaTeXTableLine}{\cpvar{fp}, $j$}
\State $\mpair{n}{m}\gets \ProcName{CvtJ2NM}(j)$;
\State \ProcName{Fprintf}(\cpvar{fp}, \lstinline|" $%d$  & $%d$  & $%d$"|, $j$, $n$, $m$); \quad // print the indices $j, n, m$ 
\State \ProcName{Fprintf}(\cpvar{fp}, \lstinline|"  &$"|);
\State \ProcName{GenSexprRadiNormCoef}(\cpvar{fp}, $\mpair{n}{m}$); \quad // print the normalization coefficient $N^m_n$
\State \ProcName{Fprintf}(\cpvar{fp}, \lstinline|"$"|); 
\State \ProcName{Fprintf}(\cpvar{fp}, \lstinline|"  &$"|);
\State \ProcName{GenSexprNonStandardZern}(\cpvar{fp}, $\mpair{n}{m}$); \quad // print $\hat{\Zern}_j(\rho,\theta)=\Radipoly{n}{m}(\rho)\Theta_m(\theta)$
\State \ProcName{Fprintf}(\cpvar{fp}, \lstinline|"$"|); 
\EndFunction
\end{algorithmic}
\end{breakablealgorithm}

\subsubsection{Generate \LaTeX{} Code for Long Table}
                           
We can set the format of the long table based on the  \lstinline|\usepackage{longtable}| in the preamble of the \LaTeX{} source file. The procedure \ProcName{GenLaTeXLongTable} in \Algr \ref{alg-genLaTeX-longtable} is used for automatically generating a long table which may span multiple pages.

\begin{breakablealgorithm}
\caption{Generate \LaTeX{} code for a long table with the data sheet of size $\scrd{n}{rows}\times \scrd{n}{cols}$ where
 $\scrd{n}{rows} = \scrd{i}{end}-\scrd{i}{begin}+1$.}
 \label{alg-genLaTeX-longtable}
\begin{algorithmic}[1]
\Require Argument \cpvar{fp} for \LaTeX{} file, integer $\scrd{i}{begin}$, integer $\scrd{i}{end}$, variable \cpvar{tab} for table information with the four members:
   \begin{itemize}
   \item $\scrd{n}{cols}$: number of attribute
   \item \cpvar{attribname}: list of attribute names
   \item \cpvar{alignctrl}: alignment information
   \item \cpvar{caption}: name of the table title
   \end{itemize}
   
\Ensure \LaTeX{} code for the long table  
\begin{equation*}
\mathscr{T}=\set{d^\alpha\in S_1\times S_2 \times \cdots \times S_{\scrd{n}{cols}}: \scrd{i}{begin}\le \alpha \le \scrd{i}{end}}
\end{equation*} 
\Function{GenLaTeXLongTable}{\cpvar{fp}, \cpvar{tab}, $\scrd{i}{begin}$, $\scrd{i}{end}$}
    \State \ProcName{SetLtabBeginCenter}(\cpvar{fp});                               
    \State \ProcName{SetLtabBeginLtable}(\cpvar{fp}, \cpvar{tab});          
    \State \ProcName{SetLtabCaption}(\cpvar{fp}, \cpvar{tab});                        
    \State \ProcName{SetLtabHline}(\cpvar{fp});                                     
    \State \ProcName{SetLtabTableHead}(\cpvar{fp}, \cpvar{tab});                 
    \State \ProcName{SetLtabHline}(\cpvar{fp});                                     
    \State \ProcName{SetLtabEndFirstHead}(\cpvar{fp});                              
    \State \ProcName{SetLtabCaptionContinue}(\cpvar{fp}, \cpvar{tab}); 
    \State \ProcName{SetLtabHline}(\cpvar{fp});                                                                        
    \State \ProcName{SetLtabTableHead}(\cpvar{fp}, \cpvar{tab});                             
    \State \ProcName{SetLtabHline}(\cpvar{fp});                                                                         
    \State \ProcName{SetLtabEndHead}(\cpvar{fp});                                   
    \State \ProcName{SetLtabHline}(\cpvar{fp});                                                                          
    \State \ProcName{SetLtabEndFoot}(\cpvar{fp});                                   
    \State \ProcName{SetLtabHline}(\cpvar{fp});                                                                         
    \State \ProcName{SetLtabEndLastFoot}(\cpvar{fp});                               
    \For{$i\in \seq{\scrd{i}{begin}, \scrd{i}{begin} + 1, \cdots, \scrd{i}{end}}$}
        \State \ProcName{GenLaTeXTableLine}(\cpvar{fp}, $i$);   // depends on concrete problem
    \EndFor
    \State \ProcName{SetLtabHline}(\cpvar{fp});                              
    \State \ProcName{SetLtabEndLtable}(\cpvar{fp});                       
    \State \ProcName{SetLtabEndCenter}(\cpvar{fp}); 
\EndFunction
\end{algorithmic}
\end{breakablealgorithm}

\subsubsection{Generate Main Body of \LaTeX{} Source File}

The main body of the \LaTeX{} source file is a necessary part of the \LaTeX{} source file, which has simple specification. The procedure \ProcName{GenLaTeXFileMainBody}
 in \Algr \ref{alg-latex-mainbody} is used for generating the main body of \LaTeX{} source file. For controlling the the format of the pdf file to be created, it is necessary for us to set the document class of the \LaTeX{} source file. We just set it with the mode "article" and select the 11pt fonts and A4 paper. Please see the procedure 
 \ProcName{SetLaTeXDocCalss} in \Algr \ref{alg-latex-doc-class}.
  
\begin{breakablealgorithm}
\caption{Generate \LaTeX{} code for the main body of the \LaTeX{} source file for generating a long table}
\label{alg-latex-mainbody}
\begin{algorithmic}[1]
\Require Argument \cpvar{fp} for \LaTeX{} file, integer $\scrd{i}{begin}$, integer $\scrd{i}{end}$
\Ensure Main body of the \LaTeX{} source file for generating a long table
\Function{GenLaTeXFileMainBody}{\cpvar{fp}, \cpvar{tab}, $\scrd{i}{begin}$, $\scrd{i}{end}$}
\State \ProcName{Fprintf}(\cpvar{fp}, \lstinline|"\\begin{document}\n\n"|);
\State \ProcName{Fprintf}(\cpvar{fp}, \lstinline|"\\maketitle\n\n"|);
\State \ProcName{Fprintf}(\cpvar{fp}, \lstinline|"\n"|);
\State \ProcName{GenLaTeXLongTable}(\cpvar{fp}, \cpvar{tab}, $\scrd{i}{begin}$, $\scrd{i}{end}$);
\State \ProcName{Fprintf}(\cpvar{fp}, \lstinline|"\n"|);
\State \ProcName{Fprintf}(\cpvar{fp}, \lstinline|"\\end{document}\n"|);
\EndFunction
\end{algorithmic}
\end{breakablealgorithm}

\begin{breakablealgorithm}
\caption{Set the document class of the \LaTeX{} source file}
\label{alg-latex-doc-class}
\begin{algorithmic}[1]
\Require Argument \cpvar{fp} for \LaTeX{} source file
\Ensure  Text line for the document class
\Function{SetLaTeXDocCalss}{\cpvar{fp}}
\State \ProcName{Fprintf}(\cpvar{fp}, \lstinline|"\\documentclass[11pt,a4paper]{article}\n\n"|);
\EndFunction
\end{algorithmic}
\end{breakablealgorithm}

\subsubsection{Generate \LaTeX{} Source File}

In the \LaTeX{} programming, it is important for us to set the preamble so as to import the macros or definitions for special symbols, mathematical environments, format specifications and so on. For our purpose, we need to import packages for mathematics, long table and paper size. Particularly, we should define new commands for printing the mathematical notations $\Zernpoly{n}{m}$ and $\Radipoly{n}{m}$. The procedure \ProcName{SetLaTeXDocPreamble} in \Algr
\ref{alg-latex-set-preamble} describes the details about setting the preamble.
  
\begin{breakablealgorithm}
\caption{Set the preamble of the \LaTeX{} source file}
\label{alg-latex-set-preamble}
\begin{algorithmic}[1]
\Require Argument \cpvar{fp} for \LaTeX{} source file, struct variable \cpvar{layout} for the layout information which includes the following members:
\begin{itemize}
\item \cpvar{orient}: orientation, it could be \lstinline|portrait| or \lstinline|landscape|
\item \cpvar{left}: distance for the left margin, say \lstinline|"2.0cm"|
\item \cpvar{right}: distance for the right margin, say \lstinline|"2.0cm"|
\item \cpvar{top}: distance for the top margin, say \lstinline|"2.0cm"|
\item \cpvar{bottom}: distance for the bottom margin, say \lstinline|"2.0cm"|
\end{itemize} 
\Ensure  Text lines for the preamble of the \LaTeX{} source file
\Function{SetLaTeXDocPreamble}{\cpvar{fp},\cpvar{layout}}
\State //Import \LaTeX{} packages required for mathematical formula, paper size and long table
\State \ProcName{Fprintf}(\cpvar{fp}, \lstinline|"\\usepackage{amsmath,amsfonts,amssymb}\n"|);
\State \ProcName{Fprintf}(\cpvar{fp}, \lstinline|"\\usepackage[%s,left=%s,right=%s,top=%s,bottom=%s]{geometry}\n"|,
	            \cpvar{layout.orient},
	            \cpvar{layout.left},
	            \cpvar{layout.right},
	            \cpvar{layout.top},
	            \cpvar{layout.bottom});
\State \ProcName{Fprintf}(\cpvar{fp}, \lstinline|"\\usepackage{longtable}\n\n"|);
\State //Set the macros for generating the symbolic expresstions for Zernike functions:
\State \ProcName{Fprintf}(\cpvar{fp}, \lstinline|"\\DeclareMathOperator{\\Zern}{Z}\n"|);
\State \ProcName{Fprintf}(\cpvar{fp}, \lstinline|"\\DeclareMathOperator{\\Radi}{R}\n"|);
\State \ProcName{Fprintf}(\cpvar{fp},\lstinline|"\\newcommand{\\Zernpoly}[2]{\\Zern_{#1}^{#2}}\n"|);
\State \ProcName{Fprintf}(\cpvar{fp},\lstinline|"\\newcommand{\\Radipoly}[2]{\\Radi_{#1}^{#2}}\n"|);
\State //Set the title and author for the pdf document to be generated 
\State \ProcName{Fprintf}(\cpvar{fp}, \lstinline|"\\title{Table of Zernike Circular Polynoms}\n"|);
\State \ProcName{Fprintf}(\cpvar{fp}, \lstinline|"\\author{...}\n"|);
\EndFunction
\end{algorithmic}
\end{breakablealgorithm}

The key contents of \LaTeX{} source file is the code in the \LaTeX{} environment \lstinline|\begin{document} ... \end{document}|, which includes the long table generated by the procedure \ProcName{GenLaTeXLtable} in \Algr \ref{alg-genLaTeX-longtable}. The procedure \ProcName{GenKeyContentsInTeXFile} in \Algr \ref{alg-latex-key-contents} shows the mechanicsm of generating the very \LaTeX{} code of interest. 

\begin{breakablealgorithm}
\caption{Generate the key contents, i.e. the \LaTeX{} long table, for the \LaTeX{} source file}
\label{alg-latex-key-contents}
\begin{algorithmic}[1]
\Require Argument \cpvar{fp} for \LaTeX source file, struct variable \cpvar{tab} with four members (viz. $\scrd{n}{cols}$, \cpvar{atribname}, \cpvar{alignctrl} and \cpvar{caption}), integer $\scrd{i}{begin}$, and integer $\scrd{i}{end}$
\Ensure The complete \LaTeX{} code for creating long table
\Function{GenKeyContentsInTeXFile}{\cpvar{fp}, \cpvar{tab}, $\scrd{i}{begin}$, $\scrd{i}{end}$}
\State	\ProcName{Fprintf}(\cpvar{fp}, \lstinline|"\\begin{document}\n\n"|);
\State	\ProcName{Fprintf}(\cpvar{fp}, \lstinline|"\\maketitle\n\n"|);
\State	\ProcName{Fprintf}(\cpvar{fp}, \lstinline|"\n"|);	
\State	\ProcName{GenLaTeXLtable}(\cpvar{fp}, \cpvar{tab}, $\scrd{i}{begin}$, $\scrd{i}{end}$);		
\State	\ProcName{Fprintf}(\cpvar{fp}, \lstinline|"\n"|);
\State  \ProcName{Fprintf}(\cpvar{fp}, \lstinline|"\\end{document}\n"|);
\EndFunction
\end{algorithmic}
\end{breakablealgorithm}

The ultimate goal of automatically generating mathematical formula of ZCP is to create a  \LaTeX{} source file which consists of the following steps: 
\begin{itemize}
\item creating an empty \LaTeX{} *.tex file with the operation mode "write";
\item setting the document class; 
\item setting the preamble; 
\item generating the key contents in of the \LaTeX{} file, and 
\item closing the *.tex properly.
\end{itemize}
The procedure   \ProcName{GenLaTeXFile} in \Algr \ref{alg-gen-latex-source-file} demonstrates the above steps clearly. 

\begin{breakablealgorithm}
\caption{Generate \LaTeX{} source file *.tex for producing the long table of Zernike circular polynomials, viz. $\scrd{\mathscr{T}}{Zernike} = \set{d^j=[j, m, n, N^m_n, \Radipoly{n}{m}(\rho)\Theta_m(\theta)]: \scrd{j}{min}\le j \le \scrd{j}{max}}$} 
\label{alg-gen-latex-source-file}
\begin{algorithmic}[1]
\Require String \cpvar{filename} with the suffix \cpvar{.tex} for \LaTeX{} source file, struct variable \cpvar{tab} with four members (viz. $\scrd{n}{cols}$, \cpvar{atribname}, \cpvar{alignctrl} and \cpvar{caption}), minimum single index $\scrd{j}{min}$, maximum single index $\scrd{j}{max}$
\Ensure \LaTeX{} source file with the name \cpvar{filename}
\Function{GenLaTeXFile}{\cpvar{filename}, \cpvar{tab}, $\scrd{j}{min}$, $\scrd{j}{max}$}
\State \cpvar{fp} $\gets$ \ProcName{Fopen}(\cpvar{filename}, \lstinline|"w"|);\quad // Create the \LaTeX{} source file with the suffix *.tex
\State \ProcName{SetDocumentCalss}(\cpvar{fp});
\State \ProcName{SetDocumentPreamble}(\cpvar{fp});
\State \ProcName{GenKeyContentsInTeXFile}(\cpvar{fp}, \cpvar{tab}, $\scrd{j}{min}$, $\scrd{j}{max}$);    
\State \ProcName{Fclose}(\cpvar{fp}); \quad // Close the \LaTeX{} source file 
\EndFunction
\end{algorithmic}
\end{breakablealgorithm}

\subsubsection{\LaTeX{} Compiling}

Generally, the \LaTeX{} source code for editing should be compiled with the terminal of Unix/Linux/Windows operating system or with an IDE such as TeXMaker or TeXStudio. Fortunately, there are some application program interface (API) for the operating system in high level programming language. For example, in C/C++, we can use the built-in function \lstinline|system(command_expr)| to compile the source file. Here the \lstinline|command_expr| stands for the  commands of \LaTeX{} compiling with the type \lstinline|const char*|. Typical implementations for \lstinline|command_expr| could be the string \lstinline|"xelatex filename.tex"| or \lstinline|"pdflatex filename.tex"|.

\section{Conclusion} \label{sec-conclusion}

For the ZCP $\Zern_j(\rho,\theta)=N^m_n\Radipoly{n}{m}(\rho)\Theta_m(\theta)$, our new theorems show that the conversion of Noll's single index $j$ and Born-Wolf's double indices $\mpair{n}{m}$ can be implemented via  \eqref{eq-nm2j}, \eqref{eq-j2n}, \eqref{eq-j2r} and \eqref{eq-j2m}. The inter-conversion is simple, complete and satisfactory. The symbolic expression of ZCP can be automatically generated with the  \ProcName{GenLaTeXFile} in \Algr \ref{alg-gen-latex-source-file}. For the convenience of theoretic analysis and engineering design, a system architecture of generating long table of mathematical expressions of ZCP is proposed with the help of meta-programming \& \LaTeX{} programming for computer-output typesetting. The value of the new paradigm for generating ZCP lies in three merits: editing mathematical expressions automatically instead of manually, avoiding potential errors by algorithms and programs that are verified, and saving the time overhead needed in manual operations.     

As a useful reference, the mathematical expressions for $\set{\Zern_j(\rho,\theta): 1\le j \le 465}$ are provided on the GitHub site, which would be sufficient for the purpose of R \& D. For the users without interest of the underlying principles and implementation details, they can just download the mathematical table of ZCP released on the GitHub web site. 

The implementation of our algorithms is based on the C programming language and the API of OS (the standard library function \lstinline|system| in the standard C). It is easy to implement the algorithms with other programming languages which can deal with strings more conveniently such as C++, Octave, MATLAB, Python and so on.
The method for automatically generating long table of mathematical expressions for ZCP can be modified slightly so as to create various long tables in optics engineering as well as in other science and technology fields, which helps to make different kinds of handbook involving massive tables. 

In the sense of project-driven science-tchnology-engineering-mathematics (STEM) education, the automatic method for generating symbolic expressions of ZCP can be used
to create a comprehensive project for training college students' ability of solving complex problem by combining multi-disciplinary knowledge and methods. 

\section*{Data Availability Statement}

The code for automatically generating the table of Zernike circular polynomials can be downloaded from the GitHub site 
\blue{\url{https://github.com/GrAbsRD/ZernikeSymbolicExpression}}.
For the readers who has no interest in the principles and implementation 
of the algorithms developed in this paper, they can just download the pdf file 
\lstinline|TableZernikePolynom-1-465.pdf| 
or the  \LaTeX{} source file 
\lstinline|TableZernikePolynom-1-465.tex|
to generate the long table of 465 Zernike circular polynomials $\Zern_j(\rho,\theta)$ such that $1=\scrd{j}{min} \le j \le \scrd{j}{max}=465$ and $\abs{m}\le \scrd{n}{max}=29$. We believe that this table will satisfies the requirements of optics design.


\begin{thebibliography}{10}

\bibitem{BornWolf-1999}
Max Born and Emil Wolf.
\newblock {\em Principles of Optics}.
\newblock Cambridge University Press, London, 7 edition, 1999.

\bibitem{Mahajan-1981}
Virendra~N. Mahajan.
\newblock Zernike annular polynomials for imaging systems with annular pupils.
\newblock {\em Journal of The Optical Society of America}, 71(1):75--85, Jan
  1981.

\bibitem{Jasssen-2010}
A.~J. E.~M. Janssen.
\newblock {Zernike circle polynomials and infinite integrals involving the
  product of Bessel functions}.
\newblock online, Jul 5 2010.
\newblock \url{arXiv:1007.0667v1} [math-ph].

\bibitem{Shakibaei-2013}
Barmak~Honarvar Shakibaei and Raveendran Paramesran.
\newblock {Recursive formula to compute Zernike radial polynomials}.
\newblock {\em Optics Letters}, 38(14):2487--2489, Jul 2013.

\bibitem{Diaz-2013}
Jos\'{e}~Antonio D\'{i}az and Virendra~N. Mahajan.
\newblock Orthonormal aberration polynomials for optical systems with circular
  and annular sector pupils.
\newblock {\em Applied Optics}, 52(6):1136--1147, Feb 2013.

\bibitem{Mathar-2008}
Richard~J. Mathar.
\newblock Zernike basis to cartesian transformations.
\newblock online, Sept. 13 2008.
\newblock \url{arXiv:0809.2368v1} [math-ph].

\bibitem{Buhren2018}
Jens B{\"u}hren.
\newblock {\em Zernike Coefficients}, pages 1945--1946.
\newblock Springer Berlin Heidelberg, Berlin, Heidelberg, 2018.

\bibitem{Berger2022-ZernikeAberr}
Christopher~George Berger.
\newblock {Zernike Aberrations}.
\newblock Optics: The Website, \url{https://opticsthewebsite.com/Zernike},
  Accessed on 08/18/2022.

\bibitem{OptikShopTest-2007}
Daniel Malacara, editor.
\newblock {\em Optical Shop Testing}.
\newblock John Wiley \& Sons Inc, New York, 3 edition, 2007.

\bibitem{Zemax-2008}
ZEMAX Development Corporation.
\newblock {\em {ZEMAX: Optical Design Program User's Guide}}, 2008.
\newblock \url{www.zemax.com}.

\bibitem{Noll-1976}
Robert~J. Noll.
\newblock Zernike polynomials and atmospheric turbulence.
\newblock {\em Journal of the Optical Society of America}, 66(3):207--211, Mar
  1976.

\bibitem{Chong-2003}
C.~W. Chong, P.~Raveendran, and R.~Mukundan.
\newblock {A comparative analysis of algorithms for fast computation of Zernike
  moments}.
\newblock {\em Pattern Recognition}, 36(3):731--742, 2003.

\bibitem{Kintner-1976}
E.~C. Kintner.
\newblock {On the Mathematical Properties of the Zernike Polynomials}.
\newblock {\em Optics Acta}, 23(8):679--680, 1976.

\bibitem{BbtZernike2022}
Hong-Yan Zhang, Yu~Zhou, and Zhi-Qiang Feng.
\newblock Balanced binary tree schemes for computing zernike radial
  polynomials, 2022.
\newblock \url{arXiv:2212.02495v2}[math.NA].

\bibitem{TAOCP-1}
Donald~E. Knuth.
\newblock {\em The Art of Computer Programming}, volume 1: Fundamental
  Algorithms.
\newblock Addison-Wesley, New York, 3 edition, 1997.

\bibitem{GrahamKnuth-1994}
Ronald Graham, Donald~E. Knuth, and Oren Patashnik.
\newblock {\em Concrete Mathematics: A Foundation for Computer Science}.
\newblock Addison-Wesley Professional, New York, 2 edition, 1994.

\bibitem{SJZhang1996}
Shan-Jie Zhang and Jian-Ming Jin.
\newblock {\em Computation of Special Functions}.
\newblock Wiley-Interscience, New York, 1996.
\newblock the Chinese version was published in 2011 by the Nanjing University
  Press in China.

\bibitem{WyantZernike}
J.~C. Wyant.
\newblock Zernike polynomials.
\newblock \url{http://wyant.optics.arizona.edu/zernikes/zernikes.htm}.

\bibitem{Mahajan-2013-vol-3}
Virendra~N. Mahajan.
\newblock {\em Optical Imaging and Aberrations, Part III: Wavefront Analysis}.
\newblock SPIE, Washington, 2013.

\end{thebibliography}
\end{document}